\RequirePackage{lineno}

\documentclass[aps,prb,twocolumn,showpacs]{revtex4}
\usepackage{hyperref}
\usepackage{graphicx}
\usepackage[utf8]{inputenc}
\DeclareGraphicsExtensions{.png,.jpg,.eps,.pdf}

\usepackage{xcolor}
\usepackage{amsmath}

\begin{document}

\title{Frustrated magnetic interactions in a cyclacene crystal}

\author{R. Ortiz$^{1,2}$, J. C. Sancho-Garc\'ia$^{2}$, J. Fern\'andez-Rossier$^{1,3}$\footnote{On leave from Departamento de F\'isica Aplicada, Universidad de Alicante, Spain} }

\affiliation{(1)Departamento de F\'{\i}sica Aplicada, Universidad de Alicante, 03690,  Sant Vicent del Raspeig, Spain
}

\affiliation{(2)Departamento de Qu\'{\i}mica F\'{\i}sica, Universidad de Alicante, 03690, Sant Vicent del Raspeig, Spain
}

\affiliation{(3) QuantaLab, International Iberian Nanotechnology Laboratory (INL), Av. Mestre Jos\'e Veiga, 4715-330 Braga, Portugal
}

\date{\today}

\begin{abstract}
We study the emergence of magnetism and its interplay with   structural properties  in a two dimensional molecular crystal of cyclacenes, using density functional  theory (DFT). Isolated cyclacenes with an even number of fused benzenes  host two  unpaired electrons in two topological protected zero modes, at  the top and bottom carbon rings that form the molecule. We show that, in the gas phase,  electron repulsion promotes an open-shell singlet with strong intramolecular antiferromagnetic exchange. 
We consider a closed packing  triangular lattice crystal phase and
we  find a strong  dependence of the band structure and magnetic interactions on the rotation angle of the cyclacenes with respect to the crystal lattice vectors.  The orientational ground state maximizes the intermolecular hybridization, yet local moments survive.  
Intermolecular exchange is  computed to be antiferromagnetic, and DFT predicts a broken symmetry  120$^\circ$  spin phase
 reflecting the frustration of the intermolecular spin coupling. 
  Thus, the cyclacene crystal realizes a bilayer of two antiferromagnetically coupled $S=1/2$ triangular lattices. Our results provide a  bottom-up route towards carbon based strongly correlated platforms in two dimensions.

\end{abstract}

\maketitle


\section{Introduction}
Magnetism and strongly correlated phases have been traditionally  alien to the realm of graphitic crystals  and polycyclic aromatic hydrocarbons (PAH). Thus,  graphite, graphene and carbon nanotubes feature wide bands  with extended states (small or vanishing density of states at the Fermi energy)  so that  the effect of Coulomb interactions in these systems are mostly negligible\cite{}.  In spite of this, strong correlations and magnetism were predicted in graphene zigzag edges\cite{nakada1996,fujita1996,JFR07} and radical PAH\cite{morita11}, but in both cases their strong reactivity is expected to compromise their chemical stability\cite{}.  

However, recent  experimental breakthroughs  have significantly changed this scenario. First, the discovery of strongly correlated electronic phases in twisted bilayer graphene\cite{cao2018correlated,cao2018unconventional} has shown that narrow bands are indeed possible in suitably designed carbon based 2D materials. Second, the on-surface assisted synthesis of multiradical PAHs, that remain stable  in ultrahigh vacuum, and the study of their electronic excitations using   scanning tunneling  microcope spectrocopy (STS) have significantly paved the way towards the fine control of their synthesis.  This includes  triangulenes\cite{pavlivcek2017synthesis,mishra2019synthesis,mishra2020,mishra2021} and the Clar's Goblet\cite{mishra19b}, among others \cite{mishra19b,mishra2020}. Therefore, using STS\cite{ortiz2020probing} to probe the  fluctuating local moments, from PAH with open-shell singlets\cite{mishra19b,mishra2020,mishra2021}, is a firmly stablished tool for this kind of systems. 

An ingredient common to 
 both  twisted bilayer and  radical PAHs is the presence of  localized electronic states  close to the Fermi energy that host  the strongly correlated electrons. This leads us to propose here
 a bottom-up route to engineer strongly correlated carbon based 2D crystals.   Our approach  is to use as building blocks radical  molecules that can assemble in such a way that intermolecular interactions lead to a weak hybridization of those molecular states hosting the unpaired electrons.   In-plane assembly of radical PAH leads, with some exceptions\cite{mishra2020}, to strong covalent bonding of the molecular states, compromising the ultimate goal of preserving the localized molecular levels.  
 
  Here we 
  consider cyclacenes with diradical character as building blocks for a molecular crystal  (see Figure 1) with a weak intermolecular interaction, so that the diradical character can be preserved in the crystal phase. These aromatic hydrocarbon nanobelts\cite{guo2021} are very appealing molecules subject of strong interest, both from theory and experiments alike. In the gas phase,  the local moments are hosted by   molecular orbitals formed by $\pi$ atomic orbitals, that are  topological symmetry protected zero modes\cite{koshino2014topological}. A natural configuration for their self-assembly is a closed packing crystal, where the separation between molecules is in the range of the interlayer distance in graphite.  We find this preserves the local moments of the isolated molecules in the crystal phase. 
Importantly, the resulting organization in this idealized cyclacene crystal  yields a network of states that does not live in a  bipartite lattice. This opens the door, as we will show next, to frustrated antiferromagnetic interactions that are known to promote quantum spin liquid states\cite{ANDERSON1973153,Balents2010spin}, and thus different from the widely studied  broken symmetry molecular magnetism\cite{gatteschi2006} 
and from the  Lieb theorem paradigm \cite{Lieb89,JFR07,ortiz19,mishra2021} .

\begin{figure}
 \centering
    \includegraphics[width=0.45\textwidth]{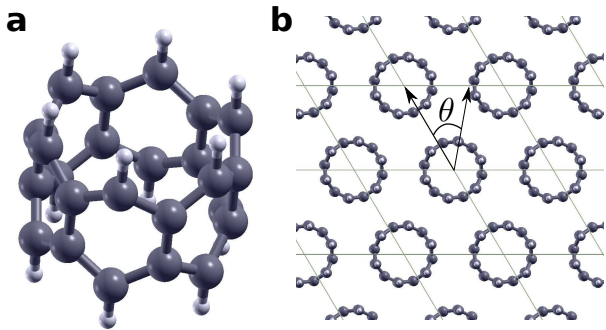}
\caption{a) Chemical structure of [6]CC and b) molecular crystal lattice of [6]crystacene showing the rotational angle $\theta$. Grey balls represent carbons and white ones represent hydrogens\cite{kokalj2003}.}
\label{fig1}
\end{figure}

Cyclacenes, hereafter [$n$]CC's, with $n$ denoting the number of fused benzene rings, can be thought  of as the shortest possible carbon nanotubes with zigzag edges. 
The electronic properties of [$n$]CC's depend  on the parity of  $n$,  showing an  odd-even  pattern\cite{choi1999structures,chen2007open,wu2016electronic}. This is confirmed at several levels of theory, where a simple one-orbital per atom tight-binding model yields two zero energy states for [even]CC's and four states split in two pairs for [odd]CC's\cite{perez2019cyclic}. This provides a simple picture to understand the di/tetraradical character obtained by more sophisticated calculations\cite{perez2018role}. On the experimental side, besides the multiple efforts to synthesize carbon nanobelts\cite{povie2017synthesis,povie2018synthesis,chisynthesis,cyclaceneDiego,jacsZhang2020,shudo2020synthesis,jacsShi2020}, the obtention of pristine cyclacenes is still missing.    

In figure \ref{fig1}a we show the  structure of [6]CC. We can think of this molecule as two carbon rings with a strong covalent coupling through the non-hydrogenated carbons (NHC). Their  tight-binding energy spectra presents two zero energy states localized on the hydrogenated carbon (HC) sites. In this work, we consider a two dimensional crystal of these  [6]CC molecules,  as if they had gone through a process of self-assembly over a surface\cite{xu1998situ,bain1988molecular,fukui2017control} (hereafter [n]crystacene, see figure \ref{fig1}b). We assume that the molecules assemble in  a triangular lattice, with the axis of the tubes perpendicular to the plane of the crystal. This idealized (monolayer) self-assembly agrees with that recently found for closely
related systems, as cycloparaphenylene nanorings\cite{perez2019structure}.  In order to characterize the orientational order of the crystal, we define the relative  angle $\theta$   formed between the line joining the geometric center of the molecule and the projection of the  hydrogenated carbon atoms on the plane containing the center. This angle is  determined by minimization of the ground state energy obtained using DFT for different magnetic   phases.

In the following we address the study of the electronic structure of the [6]crystacene, focusing on the emergence of molecular local moments, the nature of intermolecular exchange, and their interplay with the rotational angle $\theta$.

\section{Methods}   
We describe the electronic structure of cyclacenes with three different levels of theory. Individual molecules are described with a Hubbard model with a single orbital per carbon site. The model is treated both in the mean-field approximation\cite{JFR07} and using multi-configurational methods in a complete active space (CAS) with 6 single particle states and 6 electrons\cite{ortiz19,ortiz2020probing}.  The molecular crystal is treated using density functional theory(DFT) calculations, carried out with Quantum Espresso (QE) code\cite{giannozzi2009quantum,giannozzi2017advanced}. We use  a Perdew-Burke-Ernzerhof (PBE)\cite{perdew1996generalized} ultrasoft non-relativistic density functional for both carbon and hydrogen atoms. The van der Waals interactions, when considered, were implemented with the 'vdw-DF' flag\cite{thonhauser2015spin,thonhauser2007prb,berland2015van}.
The kinetic energy cut-off considered for wavefunctions was 30Ry, for the charge density and potential it was 700Ry and we employed a k-grid of 10x10x1.

We consider  5 magnetic states for the crystal: a spin-unpolarized non-magnetic state (NM),  ferromagnetic (FM), and three states with intramolecular antiferromagnetic correlations that differ by the intermolecular spin order:    AF1,  with  ferromagnetic intermolecular correlations;  AF2, a stripe phase with two antiferromagnetically coupled molecules per unit cell;  and AF120,  with non-collinear magnetic order where first neighbour spins are misaligned by $120^\circ$,  and 3 molecules per unit cell (see   figure \ref{fig2}a,b,c,d,e).

In the case of  NM, AF1 and AF2,
 we have explored the $\theta$ depedence of the ground state energy, using the
 atomic coordinates   obtained upon relaxation of the unit cell at $\theta=0^\circ$ for these three  magnetic configurations; this geometry was further employed for any angle $\theta$ and for the band structures.  
 We have also computed the FM phase for two different angles $\theta=0^\circ$ and $\theta=18^\circ$ using the same relaxed coordinates as those employed for the AF120 phase, as we will explain in the following. The results for this ferromagnetic phases are in table \ref{table1}.

In the    case of the AF120 phase, the unit cell  has 3 molecules. This makes the computational cost of the structural relaxation prohibitively large.   We use instead atomic configurations obtained with smaller unit cells and we only consider $\theta=0^\circ$. We have tried two different structures and, in both cases, the magnetic ground state  with  3 molecules in the unit cell converges to the 120 phase. Below we show results obtained using the atomic coordinates obtained from the relaxation of the
AF1 phase with $\theta=-20^\circ$,
 with the lattice constants $|\vec{a}|,|\vec{b}| = 14.29$\AA\ (which corresponds to a center-to-center intermolecular distance of 8.25\AA\ ) : the approximate position of the energy minimum in the E vs 
 $|\vec{a}|,|\vec{b}|$ curve for the frustrated AF2 phase, see suppl. mat.). No vdW interactions were considered for the AF120 phase. For the NM and AF2 phases we employed a Marzari-Vanderbilt smearing with a gaussian spreading of $10^{-4}$Ry, whilst no smearing at all was used for the AF1 or AF120 phases. \footnote{We have verified that 
  AF1 phase's energies do not depend on the smearing, that was introduced to improve  the convergence of the calculations in the neighbourhood  of $\theta=0^\circ$.}
  
  In order to characterize the magnetic state of the crystals, we use two indicators:
   \begin{equation}M_{\rm ring}^{tot}=\sum_{i\in{\rm ring}} m_i
  \end{equation}   and 
   \begin{equation}
   M_{\rm ring}^{abs}=\sum_{i\in{\rm ring}} |m_i|
   \end{equation}
   where $m_i$ is the length of the atomic  magnetic moments, as obtained from the output of our DFT calculations. The use of these two indicators 
  is motivated by  the fact that, in a given ring, there is a small residual magnetization in the non-hydrogenated carbon atoms, typical of the broken symmetry solutions in bipartite systems\cite{JFR07}.  As a reference, the  absolute magnetic moment per ring in the single molecule  mean-field calculation is $M_{\rm ring}^{abs} \sim 1.6  \mu_B$ for the FM solution, to be compared with  $M_{\rm ring}^{tot} =1 \mu_B$.

The calculation of  the atomic  magnetic moments, carried out by Quantum Espresso   by integrating the magnetic density over a sphere centered  around the atom,   turns out to be non-reliable in this system.   For instance,  let us consider the FM phase in a strongly insulating case where the total moment  per CC can be determined by counting the number of occupied bands per spin channel, and is   unambiguously determined to be  2$\mu_B$. We can now obtain  a lower limit for the magnetic moment per  hydrogenated carbon atom using the following argument.  The calculation of the magnetic density, shown in figure \ref{fig6}d, shows that non hydrogenated carbon atoms have a residual magnetization opposite to the dominant contribution of the hydrogenated carbon. We thus have $6(|m_{\rm HC}|- |m_{\rm NHC}|)=1\mu_B$, the magnetization per ring.  However, the values given by QE for this system is $6(|m_{\rm HC}|- |m_{\rm NHC}|)=0.43\mu_B$ which gives a good account of the unreliability of the integration method.
We attribute the source of  this discrepancy to the poor matching between a sphere and the $\pi$ shape electronic cloud that host the local moments (see figure 4).

For the phases with vanishing total spin per unit cell, we cannot rely on the use of the total moment to assess the error of the integration method.  For this reason, in table \ref{table1}  we  choose to normalize the values  of the atomic moments obtained by QE for all phases to the ones obtained for the FM phase with $\theta=18^\circ$.  This local moments  can be considered as an upper limit for the magnetization per ring.  We thus define:
 \begin{equation}
 \tilde{M}_{\rm ring}^{tot}=\frac{1}{M_{\rm ring}^{tot}(\rm{FM ,\theta=18^\circ})}\sum_{i\in{\rm ring}} m_i
 \label{tilde}
  \end{equation}
  and analogously for the absolute magnetization.

\section{Results and discussion}
\subsection{Single CC}
Our starting point is the single-particle spectrum of the [6]CC molecule described with the tight-binding model with one orbital per atom.  It features two zero modes whose molecular orbitals are localized at the hydrogenated carbons of the top and bottom rings.  These zero modes arise due to the combination of the chiral symmetry associated to the bipartite character of the [6]CC graph and the additional presence of the $C_N$ symmetry,  when $N$ is even, in spite of the fact that the number of sites in both sublattices is the same.

The  argument is similar to the one presented by Koshino et al\cite{koshino2014topological}.  
We use the $C_N$ symmetry to represent the Hamiltonian of the molecule. This symmetry permits to build the molecule as the repetition of $N$ blocks of 4 atoms. We label the eigenvalues of the $C_N$ symmetry as 
 $\omega^{\ell}=e^{ik\ell}$, where $\ell=0,..,N-1$. We thus have to impose  $k= \frac{2\pi}{N}m$, where $m=0,1,..,N-1$, so that  $\omega^N=1$. The Hamiltonian is thus block diagonal ($H=\sum_{k} H_0(k)$), with 
\begin{equation}
H_0(k)=t \left(\begin{array}{cccc} 
0 & 1+e^{ik} & 0 &0 \\
1+e^{-ik}& 0  & 1 &0 \\
0 & 1 & 0 &1+e^{-ik} \\
0 &0 & 1+e^{ik} &0
\end{array}\right) 
\end{equation}
where $t$ is the first neighbour hopping and the order of the basis goes from top to bottom in the unit cell.
Now, for  even $N$, we can have $k=\pi$ for $m=\frac{N}{2}$. This  leads to $1+e^{ik}=1+e^{-ik}=0$, so that specific block has 2 zero modes, localized at top and bottom rings. 
Effectively, the $k=\pi$ Hamiltonian is equivalent to  the molecule with two disconnected zero modes,  very much like the Clar's goblet\cite{ortiz19}. Note that the argument can be trivially generalized to longer tubes.

\begin{figure}[h!]
 \centering
    \includegraphics[width=0.48\textwidth]{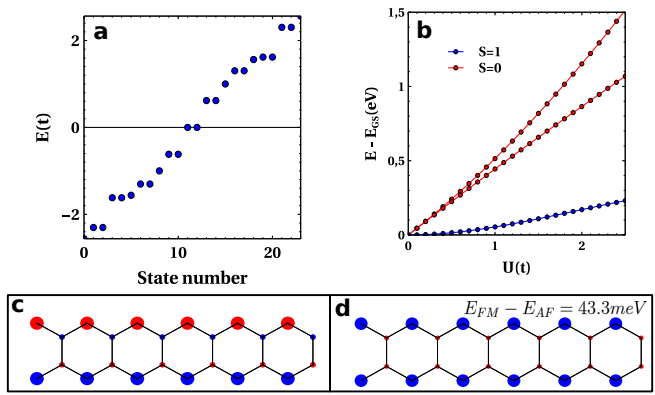}
\caption{Gas phase isolated molecule calculations. a) Single particle tight-binding spectra for [6]CC, calculated for the planar lattices of panels c) and d) with periodic boundary conditions. Note that, as we mention in the text, finite-temperature DFT calculations\cite{perez2018role} hold the diradical character of [even]CC, so ignoring the ring curvature on this matter is justified. b) Excitation energies computed with the CAS(6,6) approximation for the Hubbard model with $t=2.7eV$. The ground state has $S=0$. The $S=1$ state has $E_{S=1}=54$ meV for $U=t$. c),d)  Expectation value of   spin density $S_z(i)$, calculated in the mean-field approximation of the Hubbard model  with $U=t$ for two magnetic configurations,  with total $S_z=0$ (AF) and total $S_z=2$ (FM). The AF is the ground state with $E_{FM}-E_{AF}=43.3$ meV for $U=t$ and $t=2.7eV$ in the mean-field, not far from the CAS(6,6) result.}
\label{fig2}
\end{figure}

Our next step is to study the effect of electron-electron interaction, within the Hubbard approximation. First we carry out  a mean-field calculation for two configurations with $S_z=0$ (AF)  and $S_z=1$ (FM) .  The expectation values of the local spin density $S_z(i)$  for the ground states of the AF and FM solutions are shown in figure 2b and 2c for $U=t$.  The hydrogenated carbon atoms host a magnetic moment $g\mu_B S_z(i)$ close to $1/6$, the value expected for two unpaired electrons localized at   the zero modes. The AF solution has an energy of 43.3 meV smaller than the FM solution for $U=t$ and $t=2.7 eV$.

The mean-field results are validated by the CAS(6,6) calculations.  We systematically find that the ground state has $S=0$, complying with Lieb's theorem\cite{Lieb89}. The first excited state has $S=1$, and the energy difference grows with $U$. For $U=t$ the energy difference is 53.8 meV, close to the mean-field value.  The nature of the  intramolecular antiferromagnetic interaction is Coulomb driven exchange, very much like in bowtie diradicals\cite{ortiz19,mishra19b}. The emerging picture is therefore that the ground state is an open-shell singlet, very much like the bowtie molecule, with two unpaired electrons hosted at the zero modes.

\subsection{Atomic structure of the crystal phases}
The combination of magnetic states,  intermolecular distance and angular orientation define a huge configurational space  that can be explored only in part.  
For $\theta=0^\circ$ we find that the relaxed
 lattice constants for the NM phase were $|\vec{a}|,|\vec{b}| \approx 8.07$\AA , for the AF1 phase $|\vec{a}|,|\vec{b}| \approx 8.18$\AA , and for the AF2 phase $|\vec{a}| \approx 16.25$\AA\ and $|\vec{b}|\approx8.14$\AA . The center to center distance $d$ can be related to the [6]CC radius by $d=2R+\delta$, where $\delta\simeq 3.34$\AA\ for the AF2 phase, similar to the interlayer distance in graphite\cite{bacon1951interlayer}.

We find that, for the NM, AF1 and AF2 magnetic states, and reasonable intermolecular distances,  the  orientational ground state energy  is minimized for $\theta=0^\circ$ (see figure  \ref{fig2}f).
 Figure \ref{fig2}f  shows $E(\theta)$ for $\theta$ in a $60^\circ$ window,  on account of the $C_6$ symmetry of the system. It is apparent that
  intermolecular exchange depends strongly on $\theta$.   This figure shows that  for $\theta=0^\circ$,  AF2 has lower energy than NM and AF1 phases and for these three magnetic states $\theta=0^\circ$ is the ground state.
 
 Importantly,  we find that AF2 has lower energy than NM for all values of $\theta$,  showing that local moments survive in the crystal phase.
  In addition, we find that the lowest energy state for $\theta=0^\circ$ is the AF120 phase (table\ref{table1}), but we could not carry out a $\theta$ dependence in this case.  This type of non-collinear magnetic state is expected to arise as a broken spin symmetry solution  in triangular lattices with antiferromagnetic exchange\cite{akagi2010spin}.  We therefore find that magnetic moments persist in the crystal phase and intermolecular exchange is antiferromagnetic.    These  are the central results of  this work.
  
\begin{figure}[h!]
 \centering
    \includegraphics[width=0.40\textwidth]{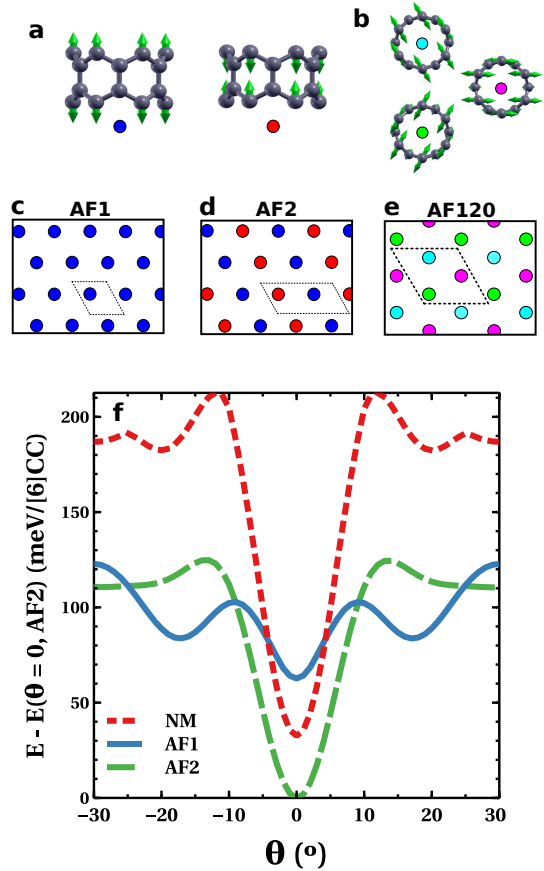}
\caption{a) Calculated local moments of the AF2 collinear phase from a side perspective when $\theta=0^\circ$ and b) AF120  phase. Hydrogens were omitted for clarity. The local moment per hydrogenated carbon is $\approx 0.04\mu_B$ in both phases.
 c), d) and e) are the schematic representations of the AF1, AF2 and AF120 phases (see text).  f) Ground state energy as a function of $\theta$,  for  the NM, AF1 and AF2 phases, referred to the value for the AF2 phase and $\theta=0^\circ$. For this calculation we employed the relaxed geometries and lattice vectors at $\theta=0^\circ$ for each magnetic configuration.}
\label{fig2}
\end{figure}
     
     \subsection{Magnetization in the crystal phases}
Our next step is to address the key question of whether the open-shell nature of the molecules  is preserved in the molecular crystal. This is assessed by computing 
 the magnitude of the local moments per carbon ring that sheds light on whether the system remains in the open-shell strong coupling limit of the single molecule, with local moments in line with those obtained in the gas phase. On the contrary, magnetic moments are quenched to some degree in the crystal phase.
    
As discussed in the methods part, it is convenient to refer our results to the case of the FM insulating crystal, for which the total magnetization is $2\mu_B$ and the magnetic moment per carbon ring is $M_{\rm ring}^{tot}=1 \mu_B$.    
  From our calculations, summarized in table \ref{table1},  we conclude that magnetic moments survive, to a large extend, in the crystal phase for the AF2, AF120 phases at $\theta=0^\circ$ and for the AF phase at $\theta=18^\circ$, with small reductions,  but they are severely quenched in the $AF1$ and $FM$ phases for $\theta=0^\circ$.  This reflects the strong interplay between orientational order and  the  survival of the magnetic moments.
  
\begin{figure}[h!]
 \centering
    \includegraphics[width=0.48\textwidth]{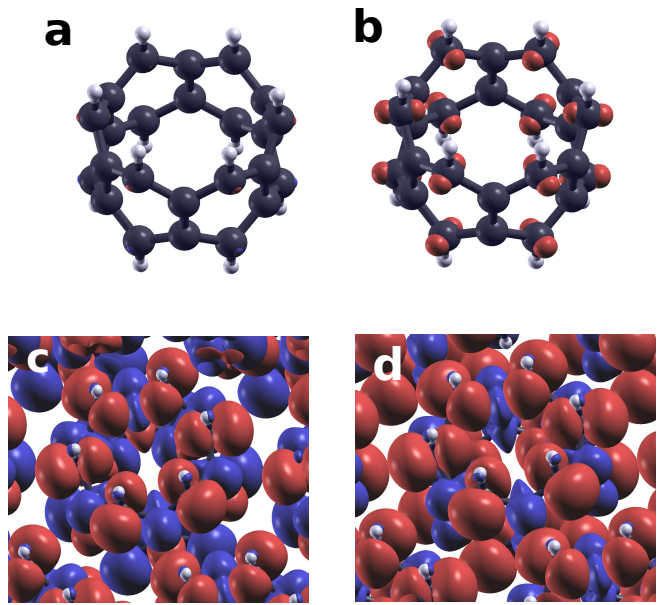}
\caption{Plot of the isosurfaces with the same isovalue\cite{kokalj2003} of the magnetic moments calculated with DFT for two different angles for the magnetic phases called in the text as a),c) AF1 with $\theta=0^\circ$ and $\theta=18^\circ$, respectively, and b),d) FM with $\theta=0^\circ$ and $\theta=18^\circ$, respectively. The color stands for the sign.}
\label{fig6}
\end{figure}

Inspection of figure \ref{fig6} clearly shows that local moments are hosted by $\pi$ orbitals.  Comparison of diffferent panels shows how the magnitude of the atomic magnetic moments is very different
in different crystal phases, in line with the results of table \ref{table1}.  It is apparent that moments are predominantly located on the hydrogenated carbon atoms, as anticipated in our single-molecule calculations with the Hubbard model. The non-hydrogenated carbons host moments opposite to the majority magnetization, as seen in  planar nanographenes \cite{JFR07}.

  \begin{table}[ht]
    \begin{tabular}{| l | l | l | l |}
    \hline
    Configuration & $\Delta E ({\rm meV})$
    & $\tilde{M}^{abs}_{\rm ring}$ & $\tilde{M}^{tot}_{\rm ring}$
%
%
    \\ \hline
    AF120 & 0.0 & 0.89 &  0.68\\ \hline
    AF2 &  6.9 & 0.92 & 0.68\\ \hline
    NM & 44.6 & 0.0 & 0.0\\ \hline
    AF1($\theta=0^\circ$) & 47.4  & 0.14 & 0.09\\ \hline
    AF1($\theta=18^\circ$) & 105.8  & 1.08 & 0.82 \\ \hline
    FM ($\theta = 0^\circ$) & 44.3 & 0.05 & 0.04\\ \hline 
    FM ($\theta=18^\circ$) &  409.2 & 1.0  & 1.0\\
    \hline
    \end{tabular}
 \caption{ Energy and magnetization for different crystacene crystal phases. $\Delta E=E-E_{AF120}$ is the energy per cyclacene referred to the energy of the $AF120$ phase, in meV. The magnetic moment per ring are relative to those computed for the  FM phase with $\theta=18^\circ$ (see Methods, equation \ref{tilde}). 
 }
\label{table1}
 \end{table}


\begin{figure}
 \centering
    \includegraphics[width=0.5\textwidth]{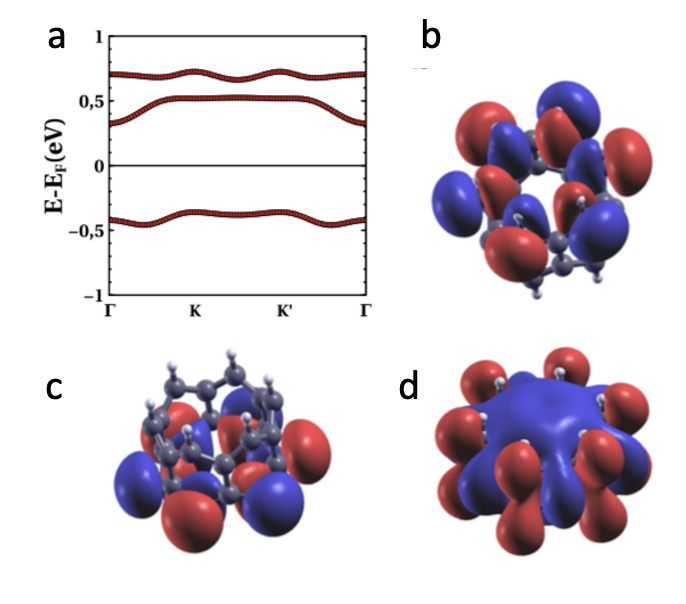}
\caption{a) Band structure for AF1 phase and $\theta=18^\circ$. b),c),d) Wannier 3D representations of the three bands in a). Panels b),c) corresponding to bottom and middle bands match the 
 zero energy states from the single-molecular system. Panel d) corresponds to the top band, which shows $\sigma$ character instead of $\pi$.}
\label{fig4}
\end{figure}

 \subsection{Energy bands}
We now look for  microscopic understanding of both   the angular dependence of the ground state energy, as well as the the differences between magnetic states. For that matter it is instructive to study the energy bands, as obtained from our DFT calculations. Naively, we expect that the bands closer to the Fermi energy are formed by the two zero modes of the diradical molecule.  These states are linear combination of $\pi$ orbitals localized along the radial direction of the molecules, which anticipate a strong dependence of the intermolecular hybridization on $\theta$.

Our energy band calculations for the AF1 phase for $\theta=18^\circ$  (figure \ref{fig4}a) show that, in addition to the two relatively narrow bands arising from the in-gap zero modes of the molecule, there is a third band slightly above, that overlaps with the states above the Fermi energy. 
The   angle $\theta=18^\circ$ corresponds  for the second minima that appears in the curve of figure \ref{fig2}f, which corresponds to a crystal geometry where carbon and hydrogen atoms avoid the alignment with those of their neighbours in order to minimize sterical interactions. In this calculation we use the relaxed unit cell with $\theta=0^\circ$,  and rotate rigidly the molecules to have  $\theta=18^\circ$. A $k$-grid of 4x4x1 was used instead since the convergence of this parameter is achieved earlier for insulators, and no smearing was used here either.
These bands feature a twofold spin degeneracy that arises in centrosymmetric antiferromagnets.

In order to confirm the origin of these bands,
 we compute the Wannier wave functions using Wannier90\cite{mostofi2014updated,marzari1997maximally,souza2001maximally}. We find three Wannier orbitals associated to the 3 narrow bands of figure \ref{fig4}a.  The ones corresponding to the first valence and conduction bands are localized in the top and bottom rings of the molecule, at  the $\pi$ orbitals of the hydrogenated  carbon atoms, as expected from our previous work for the molecular phase\cite{perez2019cyclic}.  The one corresponding to the higher energy band has a $\sigma$ character and a  smaller bandwidth thereby, on account of the smaller intermolecular hopping.

\begin{figure}
 \centering
    \includegraphics[width=0.5\textwidth]{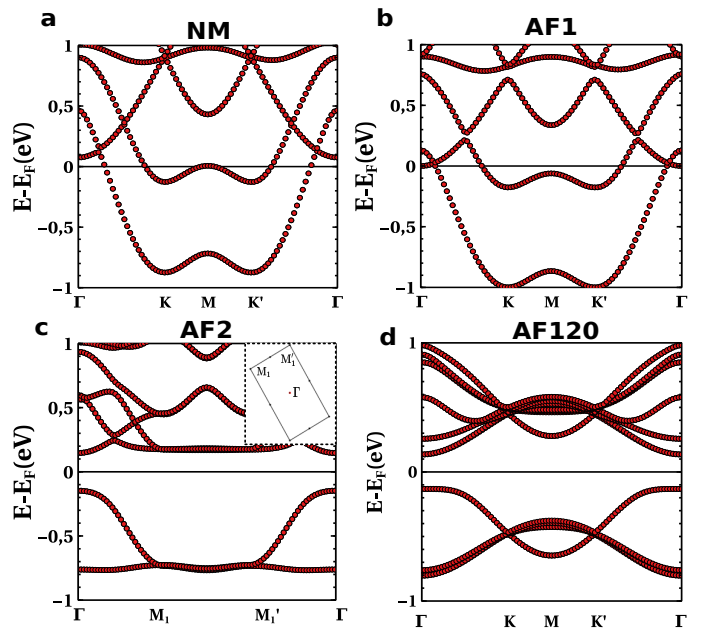}
\caption{Calculated band structure for the four magnetic configurations considered in [6]crystacene: a) NM, b) AF1, c) AF2, and d) AF120. $\theta=0^\circ$. The geometries are the same as those used for figure \ref{fig2}f. Inset in c) is the oblique lattice's first Brillouin zone with the high-symmetry points. For the collinear phases, the spin up and spin down bands are degenerate.}
\label{fig3}
\end{figure}

The width of the energy bands depends strongly on $\theta$, as can be seen comparing
\ref{fig4}a with figure \ref{fig3}b,  showing the bands of the AF1 phase for two different values of $\theta$.
We see that the bandwidth  $W$ of the valence band goes from less than 200 meV for $\theta=18^\circ$ to more than 1.1 eV for $\theta=0^\circ$. This dependence is a consequence of the 
 highly directional nature of the $\pi$ orbitals that form these bands.

In turn, the modulation of the bandwidth has consequences on the local moment formation.
The two energy scales that govern the emergence of local moments are the bandwidth $W$ of the bands at the Fermi energy and the effective Coulomb repulsion for double occupation of the molecular orbitals,  that in the Hubbard aproximation is given by $\tilde{U}=U\sum_i |\psi(i)|^2$, where $\psi(i)$ is the amplitude of the molecular orbital at atom $i$ and $U$ is the atomic Hubbard parameter,  that is in the range of 9 eV\cite{wehling2011}. Since $\psi(i)$ is equally distributed in six atoms, we estimate $\tilde{U}=U/6\simeq 1.5 eV$. This is to be compared with the bandwidth. 

In figure \ref{fig3}a,b we see how both the $\theta=0^\circ$ NM and the AF1 phases are conducting and have a Fermi surface with two pockets. In contrast,  the AF2 and AF120 phases (figure \ref{fig3}c,d), that are more stable in energy, are insulating, also for $\theta=0^\circ$. This shows the interplay between the intermolecular spin correlations and conduction.

Our results  suggest that  [6]crystacene provides a physical realization  of
two 
triangular monolayers that host local moments,  with strong  intralayer antiferromagnetic interactions, and antiferromagnetic interlayer exchange.  In the AF120 phase the quenching of the  local moments,  relative to the gas phase, is small compared to the molecular case,  
so  the system provides a physical realization of the triangular lattice bilayer Heisenberg model  \cite{singh1998}. 
. The nature of the  ground state of the S=1/2 triangular monolayer has been studied for 5 decades now \cite{ANDERSON1973153,bernu1994,capriotti99,Balents2010spin}.  
Using both quantum Monte Carlo and exact diagonalizations it was established that the ground state of the $S=1/2$ Heisenberg model with first neighbour AF coupling  features long range order with  gapless Goldstone mode excitations and depleted magnetic moments,  on account of enhanced quantum fluctuations\cite{capriotti99}.  For the   triangular bilayer\cite{singh1998} it was found that, depending on the ratio $\lambda=\frac{J}{g}$ between intralayer $J$ and interlayer $g$ exchange,  the spectrum is gapped for small $\lambda$, and gapless for $\lambda>1.4$, signalling the critical value for the transition between a quantum disordered state and   a ground state with  broken symmetry and long range order.
Interestingly, intramolecular exchange could be modified if we consider a crystal of longer cyclacenes. Therefore,  this platform could permit to study the  quantum phase transition that may occur when the ratio of intramolecular and intermolecular exchange is modified. 
  
As for the electronic properties of the broken symmetry AF120 phase predicted by our DFT results, we expect a vanishing 
Chern number \cite{takatsu2010unconventional} as well as a vanishing Berry curvature, and thereby null  Anomalous Hall effect.  {\em In contrast,} at three-quarters filling the ground state of a triangular lattice is the so-called Q-phase predicted by Martin and Batista\cite{martin2008itinerant,chern2012spontaneous}, where the spins point to the corners of a tetrahedron, and there is a  quantized anomalous Hall conductivity \cite{martin2008itinerant}. The study of this phase is out of the scope of the present paper.

\section{Conclusions}

We have studied different magnetic configurations from a molecular crystal consisting in a two-dimensional triangular array of cyclacenes.  In the gas phase, we have explored the diradical nature of the molecules using both multiconfigurational  and mean-field calculations for the Hubbard model. We have shown that the diradical nature of the molecules has a topological origin, that leads to an open-shell $S=0$ singlet formed by two antiferomagnetically coupled electrons localized at the top and bottom ring of the molecule. We have studied the fate of these local moments in the crystal phase, using   density functional theory.  We have explored both the relative orientation of the molecules with respect to the crystal lattice vectors, characterized by the angle $\theta$,  and we have considered four different magnetic states: NM and four types of magnetic order, FM, AF1, AF2, AF120.
We find a strong dependence of the electronic structure with $\theta$. 
For all angles, the AF2 magnetic solution has lower energy than the NM case, showing that the antiferromagnetically correlated local moments of the gas phase survive in the crystal.
 We find that the ground state occurs for the AF120 phase,   for $\theta=0^\circ$, which indicates antiferromagnetic intermolecular interactions.   Our results portray the crystacene crystal as a versatile platform to promote strongly correlated phases, akin to the twisted bilayer graphene.

We acknowledge J. L. Lado and A. Perez Guardiola for fruitful discussions. We ackowledge for funding support to Ministry of Science and Innovation of Spain (grant numbers PID2019-106114GB-I00 and PID2019-109539GB), Generalitat Valenciana and Fondo Social Europeo (grant number ACIF/2018/175) and 
FEDER/Junta de Andaluc\'ia Consejer\'ia de Transformaci\'on Econ\'omica, Industria,
Conocimiento y Universidades, grant PY18-4834.


\bibliographystyle{naturemag}
\bibliography{biblio}{}

\end{document}